\documentclass[11pt,a4paper,reqno]{amsart}
\usepackage[T1]{fontenc}
\usepackage[utf8]{inputenc}
\usepackage{lmodern}
\usepackage{amssymb,mathrsfs}
\usepackage{microtype,booktabs,tabularx,array}
\usepackage{xcolor}
\definecolor{linkblue}{RGB}{25,55,95}
\usepackage[colorlinks=true,linkcolor=linkblue,citecolor=linkblue,urlcolor=linkblue]{hyperref}
\hypersetup{pdftitle={Sharp Lp-Regularity for Quantum Markov Semigroups},
 pdfauthor={Lijun Wang}}
\newcommand{\Tr}{\operatorname{Tr}}
\newcommand{\Ree}{\operatorname{Re}}
\newcommand{\id}{\operatorname{id}}
\newtheorem{theorem}{Theorem}[section]
\newtheorem{proposition}[theorem]{Proposition}
\newtheorem{lemma}[theorem]{Lemma}
\newtheorem{corollary}[theorem]{Corollary}
\newtheorem*{conjecture}{Conjecture}
\theoremstyle{definition}
\newtheorem{remark}[theorem]{Remark}
\numberwithin{equation}{section}
\allowdisplaybreaks[1]
\title[Sharp \(L_p\)-Regularity for Quantum Markov Semigroups]{Sharp \(L_p\)-Regularity for Quantum Markov Semigroups}
\author{Lijun Wang}
\date{November 1, 2026}
\begin{document}

\begin{abstract}
We determine the sharp universal constants in positive-input
\(L_p\)-regularity for finite-dimensional quantum Markov
semigroups with a faithful invariant state. For \(1<p<\infty\), they are
\(p\sin(\pi/p)/(2(p-1))\) under KMS symmetry and
\(\frac{p}{2(p-1)}\tan(\frac{\pi}{2}\min\{1/p,1-1/p\})\)
without a symmetry assumption, with the endpoint
values at \(p=1\) given by \(\pi/2\) and \(\pi/4\), respectively.
\end{abstract}
\maketitle
\enlargethispage{5pt}

\section{Introduction}\label{sec:introduction}

In the seminal work \cite{Gross75}, Gross introduced a logarithmic Sobolev inequality (LSI) for Gaussian measures as an equivalent reformulation of Nelson's
hypercontractivity (HC) of the Ornstein--Uhlenbeck semigroup \cite{Nelson73}.
Gross's formulation in terms of semigroup generators provides a framework
for studying LSI and HC beyond the Gaussian setting.
Over the past decades, LSI and HC have been extensively studied
for diffusion semigroups on Riemannian manifolds
\cite{BE85,BGL14}, finite Markov chains \cite{DSC96}, and
Wiener measures on path spaces over Riemannian manifolds
\cite{Hsu97}.
Their quantum extensions include noncommutative $L_p$
formulations \cite{OZ99,KT13} and complete modified
logarithmic Sobolev inequalities \cite{GJLL25}, with
applications in quantum information theory.

For finite Markov chains, Diaconis and Saloff-Coste used LSI and HC
to obtain quantitative bounds on entropy decay and convergence to equilibrium
\cite[Theorems~3.5--3.7]{DSC96}.
For primitive quantum Markov semigroups (QMS), Kastoryano and Temme
showed how an energy comparison at $p=1$ transfers a quadratic LSI
to a modified logarithmic Sobolev inequality (MLSI)
\cite[Proposition~13 and Lemma~23]{KT13}.
They also derived exponential decay of relative entropy and,
through the quantum Pinsker inequality, trace-norm mixing bounds
\cite[Lemma~21 and Theorem~22]{KT13}.
These bounds control the error in dissipative state preparation
\cite[Theorem 3.4 and Proposition 3.5]{Watrous18}.

In the classical setting, such energy comparisons underlie
Gross's passage from LSI to HC, transferring the quadratic estimate
to other exponents.
For positive inputs, differentiating an $L_{p(t)}$ norm along the
semigroup, with $p(t)>1$ increasing in time, produces both an entropy
term and the Dirichlet form at the current exponent.
Applying the quadratic LSI to the $p(t)/2$-power of the evolved
function bounds the entropy by a quadratic energy.
Comparing this energy with the energy in the norm derivative then
allows the exponent growth to be chosen so that the norm is
nonincreasing \cite[Lemma~2.6 and Theorem~3.5]{DSC96}.
Such comparisons are formalized by $L_p$-regularity conditions.

For a reversible classical Markov semigroup $P_t=e^{tL}$ on a finite
probability space $(\Omega,\mu)$ with invariant measure $\mu$,
we define the $L_p$ Dirichlet form by
\[
 \mathcal E_p^L(f)
 =-\frac{p}{2(p-1)}\int_\Omega f^{p-1}Lf\,d\mu,
 \qquad f>0,\quad 1<p<\infty.
\]

The $L_p$--$L_2$ Dirichlet form comparison follows from the classical
Stroock--Varopoulos inequality \cite[Section~2]{MOS13}:
\[
 \mathcal E_p^L(f)
 \ge\frac2p\mathcal E_2^L(f^{p/2}),
 \qquad f>0,\quad 1<p<\infty.
\]

To extend Gross's argument to quantum semigroups, Olkiewicz
and Zegarlinski formulated an $L_p$-regularity condition in
weighted noncommutative $L_p$ spaces
\cite[Definition~3.6 and Theorem~3.8]{OZ99}.
In this work, we study the optimal constants in positive-input energy
comparisons under different symmetry assumptions.

To formulate this problem in finite dimensions, let
$T_t=e^{t\mathcal L}$ be a QMS with faithful invariant state $\sigma$.
For $1\le p<\infty$, the weighted $L_p$ norm is
$\|X\|_{p,\sigma}=\|\sigma^{\frac{1}{2p}}X\sigma^{\frac{1}{2p}}\|_p$,
where $\|\cdot\|_p$ is the Schatten $p$-norm.
The Dirichlet forms measure the dissipation of these norms along $T_t$.
For $f>0$ and $1<p<\infty$, we use the normalization
\[
 \mathcal E_p^{\mathcal L}(f)
 =-\frac1{2(p-1)}
   \left.\frac{d}{dt}\|T_tf\|_{p,\sigma}^p\right|_{t=0^+}.
\]
The endpoint $\mathcal E_1^{\mathcal L}$ is defined by the continuous
limit on each fixed $f>0$.

For $f>0$, the $L_p$-regularity comparisons use the power transform
to match the homogeneity of the $L_p$ and quadratic forms:
\[
 I_{2,p}(f)=\sigma^{-\frac{1}{4}}
 (\sigma^{\frac{1}{2p}}f\sigma^{\frac{1}{2p}})^{\frac{p}{2}}\sigma^{-\frac{1}{4}},
 \qquad 1\le p<\infty.
\]
Kastoryano and Temme formulate weak and strong $L_p$-regularity
for self-adjoint observables \cite[Definition~9]{KT13}.
Here we consider their restrictions to $f>0$, with comparison
constants $w(p)$ and $2/p$, respectively.
\begin{equation}\label{eq:strong-regularity}
 \begin{aligned}
 \mathcal E_p^{\mathcal L}(f)
 &\ge w(p)\mathcal E_2^{\mathcal L}(I_{2,p}f)
 &&\text{(weak)},\\
 \mathcal E_p^{\mathcal L}(f)
 &\ge\frac2p\mathcal E_2^{\mathcal L}(I_{2,p}f)
 &&\text{(strong)},
 \end{aligned}
 \qquad f>0,\quad 1\le p<\infty,
\end{equation}
where the weak threshold is\footnote{For
$p\ge2$, Definition~9 of the fixed arXiv version of \cite{KT13} prints
$p-1$, whereas its Lemma~18 gives $1/(p-1)$ in the same $L_p$ Dirichlet
form normalization. We use the latter convention.}
\begin{equation}\label{eq:weak-threshold}
 w(p)=
 \begin{cases}
  1,&1\le p\le2,\\
  (p-1)^{-1},&2\le p<\infty,
 \end{cases}
\end{equation}
They proposed the following two-part conjecture
\cite[Conjecture~19]{KT13}.
\begin{conjecture}[Kastoryano--Temme]
Let $\mathcal L$ generate a primitive QMS.
\begin{enumerate}
\item[(i)] $\mathcal L$ is weakly $L_p$-regular.
\item[(ii)] If $\mathcal L$ is reversible, then it is strongly
$L_p$-regular.
\end{enumerate}
\end{conjecture}
Here reversibility means KMS symmetry, which is weaker than GNS symmetry.

Cubitt, Kastoryano, Montanaro and Temme established a quantum
Stroock--Varopoulos inequality in the tracial setting of unital,
reversible, primitive quantum channels and connected it with
$L_p$-regularity
\cite[Lemma~9, Corollary~10 and the subsequent discussion]{CKMT15}.
Bardet established the strong comparison under GNS symmetry
\cite[Theorem 4.1]{Bardet17}.
Beigi, Datta and Rouz\'e obtained a weighted two-exponent comparison
under the same symmetry, including exponents below one
\cite[Theorem 14]{BDR20}.

Failure of strong regularity without reversibility was already noted in
\cite{KT13}. Becker, Rouz\'e and Salzmann gave a KMS-symmetric qubit
counterexample at the entropy endpoint
\cite[Proposition~2.10]{BRS26}.

These results leave open the optimal $L_p$-regularity constants in
the general and KMS classes. In this work, we determine these universal constants
for every finite $p\ge1$.

\begin{samepage}
\begin{theorem}\label{thm:optimal-class}
Let \(\mathcal H\) be a finite-dimensional Hilbert space and \(T_t=e^{t\mathcal L}\) be a
QMS on \(B(\mathcal H)\) with faithful invariant state \(\sigma\).
For every \(1\le p<\infty\) and \(f>0\),
\[
 \mathcal E_p^{\mathcal L}(f)
 \ge\kappa_{\mathrm G}(p)\,
       \mathcal E_2^{\mathcal L}(I_{2,p}f),
\]
where
\begin{equation}\label{eq:coefficients}
 \kappa_{\mathrm G}(p)
 =\frac{p}{2(p-1)}
 \tan\!\left(\frac{\pi}{2}\min\{1/p,1-1/p\}\right),
 \qquad 1<p<\infty,
\end{equation}
and \(\kappa_{\mathrm G}(1)=\pi/4\).
If, in addition, \(\mathcal L\) is KMS symmetric, then
\[
 \mathcal E_p^{\mathcal L}(f)
 \ge\kappa_{\mathrm K}(p)\,
       \mathcal E_2^{\mathcal L}(I_{2,p}f),
\]
where
\begin{equation}\label{eq:coefficients-kms}
 \kappa_{\mathrm K}(p)
 =\frac{p\sin(\pi/p)}{2(p-1)},
 \qquad 1<p<\infty,
\end{equation}
and \(\kappa_{\mathrm K}(1)=\pi/2\). Both constants $\kappa_{\mathrm G}$ and $\kappa_{\mathrm K}$ are sharp.
\end{theorem}
\end{samepage}

For primitive QMS satisfying GNS symmetry,
\cite[Theorem 14]{BDR20} gives the comparison with constant
\(2/p\) for \(1<p\le2\).
Appendix~\ref{sec:benchmarks} extends this bound to all finite \(p\ge1\)
without assuming primitivity and establishes sharpness.
These optimal constants and the weak threshold \(w(p)\) satisfy,
for every finite \(p\ge1\), \(p\ne2\),
\[
 \kappa_{\mathrm G}(p)<w(p)<\kappa_{\mathrm K}(p)<2/p;
\]
at \(p=2\), all four constants are \(1\).

The strict gaps therefore yield, for every finite \(p\ge1\), \(p\ne2\), a primitive general QMS violating weak
regularity and a primitive KMS-symmetric QMS violating strong regularity,
in both cases on a strictly positive input. Thus both positive-input
assertions of Conjecture~19 in \cite{KT13} are false, regardless of
the high-exponent convention.

For the functional-inequality consequence, set
$A_p(f)=(\sigma^{\frac{1}{2p}}f\sigma^{\frac{1}{2p}})^p$
for $1\le p<\infty$ and $f>0$, and define
\[
 \operatorname{Ent}_{p,\sigma}(f)
 =\frac1p\Tr\!\left[A_p(f)
       \bigl(\log A_p(f)-\log((\Tr A_p(f))\sigma)\bigr)\right].
\]
The $L_p$-logarithmic Sobolev constant $\alpha_p$ of $\mathcal L$ is the
largest constant such that
$\mathcal E_p^{\mathcal L}(f)\ge
\alpha_p\operatorname{Ent}_{p,\sigma}(f)$ for all $f>0$
\cite[Definition~11]{KT13}.
The standard regularity-to-LSI and hypercontractivity transfers
\cite[Proposition 13 and Theorem 15]{KT13},
\cite[Theorem 11]{BDR20} in the reversible setting, give the following
quantitative consequence of Theorem~\ref{thm:optimal-class}.

\begin{corollary}\label{cor:functional-inequalities}
Let $T_t=e^{t\mathcal L}$ be as in Theorem~\ref{thm:optimal-class}.
Take $\kappa=\kappa_{\mathrm G}$, or $\kappa=\kappa_{\mathrm K}$ if
$\mathcal L$ is KMS symmetric. Suppose that its $L_2$-logarithmic Sobolev constant 
$\alpha_2>0$. Then, for every finite $p\ge1$,
\[
 \alpha_p\ge\frac{p\kappa(p)}2\alpha_2.
\]
In particular,
\[
 \alpha_p\ge
 \begin{cases}
  (\pi/4)\alpha_2,&\text{in the KMS-symmetric case},\\
  (\pi/8)\alpha_2,&\text{in the general case}.
 \end{cases}
\]
Moreover, for every $f>0$ and $1<p_0\le q<\infty$,
\begin{equation}\label{eq:hc-integral}
 \|T_tf\|_{q,\sigma}\le\|f\|_{p_0,\sigma}
 \quad\text{whenever}\quad
 t\ge\frac1{\alpha_2}\int_{p_0}^q
       \frac{ds}{s(s-1)\kappa(s)}.
\end{equation}
\end{corollary}

Combining this corollary with the converse implication from
hypercontractivity to LSI gives the equivalence
\begin{equation}\label{eq:lsi-hc-equivalence}
 \alpha_2>0
 \quad\Longleftrightarrow\quad
 \begin{gathered}
  \text{there exists }a>0\text{ such that}\\
  \|T_tf\|_{1+e^{2at},\sigma}\le\|f\|_{2,\sigma}
  \qquad(t\ge0,\ f>0).
 \end{gathered}
\end{equation}
Indeed, using $\kappa_{\mathrm G}(s)\ge\pi/(4s)$ and
$\kappa_{\mathrm K}(s)\ge\pi/(2s)$ in \eqref{eq:hc-integral}, one may
take $a=(\pi/8)\alpha_2$ in general and $a=(\pi/4)\alpha_2$ in the
KMS-symmetric case. Conversely, the HC estimate with
$r(t)=1+e^{2at}$ gives, by differentiation at $t=0^+$,
\[
 0\ge
 \left.\frac{d}{dt}\log\|T_tf\|_{r(t),\sigma}\right|_{t=0^+}
 =\frac{a\operatorname{Ent}_{2,\sigma}(f)
         -\mathcal E_2^{\mathcal L}(f)}{\|f\|_{2,\sigma}^{2}},
 \qquad f>0,
\]
and hence $\alpha_2\ge a$.

Section~\ref{sec:preliminaries} recalls the $L_p$ Dirichlet form and related conventions.
Section~\ref{sec:comparison-bounds} uses a common Poisson representation
to prove the finite-channel bound in Cayley form and derive the
$L_p$-regularity bounds for generators by taking a small-time limit.
In Section~\ref{sec:sharpness}, the matching
constructions and primitive
approximation establish the optimality statements.

{\bf \noindent Acknowledgments.} The author thanks his advisor Li Gao for helpful discussions and valuable suggestions.
The author is supported in part by the National Natural
Science Foundation of China (grant no. 12401163) and the Department of
Science and Technology of Hubei Province (project nos. 2025EHA041 and
2025AFA044).
The author used ChatGPT to discuss proof ideas and technical
details. All mathematical arguments were
checked by the author, who takes full responsibility for the content of
this paper.

\section{Weighted \texorpdfstring{$L_p$}{Lp} Dirichlet forms and adjoints}
\label{sec:preliminaries}

Let $\mathcal H$ be a finite-dimensional Hilbert space and
write $B(\mathcal H)$ for its
operator algebra, with identity $I$. Throughout, $\Tr$ is the ordinary
trace and logarithms are natural. We call an operator $\sigma>0$ with
$\Tr\sigma=1$ a faithful state.
For $1\le p<\infty$, let $\|X\|_p=(\Tr|X|^p)^{1/p}$ be the
Schatten norm for the ordinary trace, and write $\|\cdot\|_\infty$
for the operator norm.

For any faithful state $\sigma$, the associated weighted $L_p$ norm
is defined by
\[
 \|X\|_{p,\sigma}
 =\bigl\|\sigma^{1/(2p)}X\sigma^{1/(2p)}\bigr\|_p.
\]
At $p=2$, the weighted norm is induced by the KMS inner product
\[
 \langle X,Y\rangle_\sigma
 =\Tr\bigl(\sigma^{1/2}X^*\sigma^{1/2}Y\bigr).
\]
For $z\in\mathbb C$, define
\[
 \Gamma_\sigma^z(X)=\sigma^{z/2}X\sigma^{z/2},
 \qquad \Gamma_\sigma=\Gamma_\sigma^1.
\]
Thus $\|X\|_{p,\sigma}=\|\Gamma_\sigma^{1/p}(X)\|_p$.
The map $\Gamma_\sigma^{1/2}$ identifies the KMS inner product with the
Hilbert--Schmidt inner product $\Tr(X^*Y)$.

All powers of a positive definite matrix use its self-adjoint logarithm.
For $f>0$ and finite $r,p>0$, define the power operators
\[
 I_{r,p}(f)=\Gamma_\sigma^{-1/r}
       \bigl[(\Gamma_\sigma^{1/p}f)^{p/r}\bigr].
\]
These power operators preserve positive definiteness and satisfy
$I_{r,s}I_{s,p}=I_{r,p}$.
For $1\le p<\infty$, they also satisfy
\[
 I_{p,p}f=f,\qquad
 \|I_{2,p}f\|_{2,\sigma}^{2}=\|f\|_{p,\sigma}^{p}.
\]

We first consider a quantum channel in the Heisenberg picture, that is,
a unital completely positive (UCP) map
$\Phi:B(\mathcal H)\to B(\mathcal H)$.
Its trace adjoint $\Phi_*$ is determined by
\[
 \Tr\bigl(\Phi_*(A)X\bigr)=\Tr\bigl(A\Phi(X)\bigr).
\]
The map $\Phi$ acts on observables, while $\Phi_*$ describes state
evolution in the Schr\"odinger picture.
If $\Phi_*(\sigma)=\sigma$, its KMS adjoint $\Phi^\sharp$ is defined by
\[
 \langle X,\Phi Y\rangle_\sigma
 =\langle\Phi^\sharp X,Y\rangle_\sigma,
 \qquad X,Y\in B(\mathcal H).
\]
Equivalently,
\[
 \Phi^\sharp=\Gamma_\sigma^{-1}\Phi_*\Gamma_\sigma.
\]
It is again UCP and preserves $\sigma$.
We call $\Phi$ KMS symmetric if $\Phi=\Phi^\sharp$, and GNS symmetric
if it is self-adjoint for the inner product
$(X,Y)\mapsto\Tr(\sigma X^*Y)$.

A quantum Markov semigroup (QMS) is a norm-continuous semigroup of
UCP maps on $B(\mathcal H)$. We write $T_t=e^{t\mathcal L}$. The trace adjoint
$\mathcal L_*$ generates the Schr\"odinger-picture semigroup $T_{t,*}$
and satisfies
\[
 \Tr\bigl(\mathcal L_*(A)X\bigr)
 =\Tr\bigl(A\mathcal L(X)\bigr).
\]
For the semigroup statements below, $\sigma$ is a faithful invariant
state, so $T_{t,*}(\sigma)=\sigma$ for every $t\ge0$.
The maps
$T_t^\sharp=\Gamma_\sigma^{-1}T_{t,*}\Gamma_\sigma$ also form a QMS
with invariant state $\sigma$, whose generator satisfies
\begin{equation}
 \mathcal L^\sharp
 =\Gamma_\sigma^{-1}\mathcal L_*\Gamma_\sigma,
 \qquad
 (\mathcal L^\sharp)_*
 =\Gamma_\sigma\mathcal L\Gamma_\sigma^{-1}.
 \label{eq:adjoints}
\end{equation}
Thus $\langle X,\mathcal LY\rangle_\sigma
=\langle\mathcal L^\sharp X,Y\rangle_\sigma$.

A QMS is KMS symmetric (respectively, GNS symmetric) when each $T_t$
has that property. Equivalently, $\mathcal L$ is self-adjoint for the
corresponding inner product. In the KMS case, this means
$\mathcal L=\mathcal L^\sharp$.

For $f>0$, $1<p<\infty$, and $p'=p/(p-1)$, we define the $L_p$ Dirichlet form
associated with $\mathcal L$ by
\begin{equation}
 \mathcal E_p^{\mathcal L}(f)
 =-\frac{p}{2(p-1)}\Ree\langle I_{p',p}(f),\mathcal Lf\rangle_\sigma
 =-\frac{p}{2(p-1)}\Tr\left[
   (\Gamma_\sigma^{1/p}f)^{p-1}\Gamma_\sigma^{1/p}(\mathcal Lf)
   \right].
 \label{eq:dirichlet}
\end{equation}
In particular,
$\mathcal E_2^{\mathcal L}(X)=-\Ree\langle X,\mathcal LX\rangle_\sigma$;
the form is defined for every self-adjoint $X$.

For positive definite matrices $A$ and $B$ with equal trace, write
$D(A\Vert B)=\Tr[A(\log A-\log B)]$.
For any $A>0$, let
\[
 \mathcal I_{\mathcal L,\sigma}(A)
 =-\Tr\bigl[\mathcal L_*(A)(\log A-\log\sigma)\bigr].
\]
For a faithful state $\rho$, $\mathcal I_{\mathcal L,\sigma}(\rho)$
is the entropy production of $\mathcal L_*$ at $\rho$. For every $A>0$,
\[
 \mathcal I_{\mathcal L,\sigma}(A)
 =-\left.\frac{d}{dt}
 D\bigl(T_{t,*}A\Vert(\Tr A)\sigma\bigr)\right|_{t=0}.
\]
The following proposition relates the $p=1$ limit of the weighted
$L_p$ Dirichlet forms to the entropy production of $(\mathcal L^\sharp)_*$.
\begin{proposition}
\label{prop:endpoint}
For every $f>0$, the limit of \eqref{eq:dirichlet} as $p\to1^+$
exists and is given by
\begin{equation}
 \mathcal E_1^{\mathcal L}(f)
 =-\frac12\Tr\bigl[(\mathcal L^\sharp)_*(\Gamma_\sigma(f))
   \bigl(\log(\Gamma_\sigma(f))-\log\sigma\bigr)\bigr]
 =\frac12\mathcal I_{\mathcal L^\sharp,\sigma}(\Gamma_\sigma(f)).
 \label{eq:energy-one}
\end{equation}
\end{proposition}

\begin{proof}
In our notation, \cite[Proposition~8(2)]{KT13} gives
\[
 \lim_{p\to1^+}\mathcal E_p^{\mathcal L}(f)
 =-\frac12\Tr\bigl[\Gamma_\sigma(\mathcal Lf)
   \bigl(\log(\Gamma_\sigma(f))-\log\sigma\bigr)\bigr].
\]
The calculation there uses invariance of the reference state but not
primitivity. By \eqref{eq:adjoints},
\[
 (\mathcal L^\sharp)_*(\Gamma_\sigma(f))
 =\Gamma_\sigma\mathcal L\Gamma_\sigma^{-1}(\Gamma_\sigma(f))
 =\Gamma_\sigma(\mathcal Lf).
\]
Substitution into the limit formula and the definition of
$\mathcal I_{\mathcal L^\sharp,\sigma}$ give \eqref{eq:energy-one}.
\end{proof}

Conjugate exponents are related by passing to the KMS adjoint.
In the KMS-symmetric case, the first identity below appears in
\cite[Proposition~8(i)]{BDR20}, with a different $L_p$ Dirichlet form normalization.
\begin{proposition}
\label{prop:duality}
For every $1<p<\infty$ and $f>0$,
\begin{equation}
 (p-1)\mathcal E_p^{\mathcal L}(f)
 =\mathcal E_{p'}^{\mathcal L^\sharp}(I_{p',p}f).
 \label{eq:energy-duality}
\end{equation}
Moreover, for every self-adjoint $X$,
\begin{equation}
 \mathcal E_2^{\mathcal L^\sharp}(X)
 =\mathcal E_2^{\mathcal L}(X).
 \label{eq:quadratic-adjoint}
\end{equation}
\end{proposition}

\begin{proof}
The composition rule gives $I_{p,p'}(I_{p',p}f)=f$.
Since $p'/(p'-1)=p$, the adjoint identity yields
\[
 \begin{aligned}
 \mathcal E_{p'}^{\mathcal L^\sharp}(I_{p',p}f)
 &=-\frac p2\Ree\langle f,\mathcal L^\sharp(I_{p',p}f)\rangle_\sigma\\
 &=-\frac p2\Ree\langle I_{p',p}f,\mathcal Lf\rangle_\sigma
  =(p-1)\mathcal E_p^{\mathcal L}(f).
 \end{aligned}
\]
Taking the real part of the KMS adjoint identity with both arguments
equal to $X$ proves \eqref{eq:quadratic-adjoint}.
\end{proof}

For a linear map $M:B(\mathcal H)\to B(\mathcal H)$, $A>0$, and
$z\in\mathbb C$, define the analytic trace function
\begin{equation}
 F_{M,A}(z)
 =\Tr\left[A^{1-z}\Gamma_\sigma^z
       M\Gamma_\sigma^{-z}(A^z)\right].
 \label{eq:channel-function}
\end{equation}
For $g>0$ and semigroup maps $T_t$, the function
$F_{T_t,g^2}(s/2)$, $0\le s\le2$, is the trace functional used in
\cite[Lemma~18]{KT13}. For a UCP map $\Phi$
preserving $\sigma$ and a faithful state $\rho$, trace duality
identifies $F_{\Phi,\rho}$ with the scalar function in
\cite[proof of Proposition~3.1]{GW26conditional} when its two states
coincide. For a QMS generator $\mathcal L$, set
\begin{equation}
 D_A^{\mathcal L}(z)=-F_{\mathcal L,A}(z).
 \label{eq:analytic-energy}
\end{equation}
For fixed $M,A,\sigma$, the function $F_{M,A}$ is entire, since
all matrix powers in \eqref{eq:channel-function} are entire.
It is also bounded on the closed strip $0\le\Ree z\le1$.
Indeed, write $z=\alpha+iy$ with $0\le\alpha\le1$.
The imaginary powers of $A$ and $\sigma$ are unitary, while
their real powers range over compact sets. Thus all matrix
factors in \eqref{eq:channel-function} are uniformly bounded
in $\alpha,y$; continuity of $M$ and the trace proves the claim.
By \eqref{eq:analytic-energy}, the same properties hold for
$D_A^{\mathcal L}$.

The following lemma relates $D_A^{\mathcal L}$ to the weighted
$L_p$ Dirichlet forms and their $p=1$ limit.

\begin{lemma}
\label{lem:energy-coordinates}
Let $\mathcal L$ generate a QMS with faithful invariant state $\sigma$.
For every $A>0$, $a>0$, and $z\in\mathbb C$,
$D_{aA}^{\mathcal L}(z)=aD_A^{\mathcal L}(z)$ and
\begin{equation}\label{eq:generator-reflection}
 D_A^{\mathcal L}(z)=D_A^{\mathcal L^\sharp}(1-z).
\end{equation}
For $1<p<\infty$ and $f>0$, put $A_p=(\Gamma_\sigma^{1/p}f)^p$.
Then
\begin{equation}
 \begin{split}
 D_{A_p}^{\mathcal L}(1/p)
   &=\frac{2(p-1)}p\mathcal E_p^{\mathcal L}(f),\\
 D_{A_p}^{\mathcal L}(1/2)
   &=\mathcal E_2^{\mathcal L}(I_{2,p}f).
 \end{split}
 \label{eq:energy-transport}
\end{equation}
The second identity also holds for $p=1$, with $A_1=\Gamma_\sigma(f)$.
The endpoint derivative satisfies
\begin{equation}\label{eq:generator-endpoint}
 -\tfrac12(D_A^{\mathcal L})'(1)
 =\mathcal E_1^{\mathcal L}(\Gamma_\sigma^{-1}A).
\end{equation}
\end{lemma}

\begin{proof}
Since $(aA)^z=a^zA^z$, scalar homogeneity follows from
$a^{1-z}a^z=a$. By \eqref{eq:adjoints}, cyclicity of the trace,
and the definition of the trace adjoint,
\[
 \begin{aligned}
 D_A^{\mathcal L^\sharp}(1-z)
 &=-\Tr\!\left[A^z\Gamma_\sigma^{-z}\mathcal L_*
                  \Gamma_\sigma^z(A^{1-z})\right]\\
 &=-\Tr\!\left[\Gamma_\sigma^{-z}(A^z)
                  \mathcal L_*(\Gamma_\sigma^z(A^{1-z}))\right]\\
 &=-\Tr\!\left[\Gamma_\sigma^z(A^{1-z})
                  \mathcal L(\Gamma_\sigma^{-z}(A^z))\right]
  =D_A^{\mathcal L}(z).
 \end{aligned}
\]

Since $A_p^{1/p}=\Gamma_\sigma^{1/p}f$,
\[
 \begin{aligned}
 D_{A_p}^{\mathcal L}(1/p)
 &=-\Tr\!\left[(\Gamma_\sigma^{1/p}f)^{p-1}
                  \Gamma_\sigma^{1/p}(\mathcal Lf)\right]
  =\frac{2(p-1)}p\mathcal E_p^{\mathcal L}(f).
 \end{aligned}
\]
For arbitrary $A>0$, set $g=\Gamma_\sigma^{-1/2}(A^{1/2})$.
Since $g$ and $\mathcal Lg$ are self-adjoint,
\[
 \begin{aligned}
 D_A^{\mathcal L}(1/2)
 &=-\Tr\!\left[A^{1/2}\Gamma_\sigma^{1/2}(\mathcal Lg)\right]\\
 &=-\Tr\!\left[\Gamma_\sigma(g)\mathcal Lg\right]
  =-\Ree\langle g,\mathcal Lg\rangle_\sigma
  =\mathcal E_2^{\mathcal L}(g).
 \end{aligned}
\]
Taking $A=A_p$ gives $g=I_{2,p}f$, including at $p=1$ with
$A_1=\Gamma_\sigma(f)$, and proves the midpoint identity in
\eqref{eq:energy-transport}.

For the endpoint, fix $A>0$ and write
$B=\Gamma_\sigma\mathcal L\Gamma_\sigma^{-1}(A)
  =(\mathcal L^\sharp)_*(A)$.
Invariance gives, for every $X$,
\[
 \Tr[\Gamma_\sigma\mathcal L(X)]
 =\Tr[\sigma\mathcal L(X)]
 =\Tr[\mathcal L_*(\sigma)X]=0.
\]
Differentiating \eqref{eq:analytic-energy} at $z=1$ therefore yields
\[
 \begin{aligned}
 (D_A^{\mathcal L})'(1)
 &=\Tr[B\log A]
   -\frac12\Tr[(\log\sigma)B+B\log\sigma]\\
 &\quad-\underbrace{\Tr\!\left[\Gamma_\sigma\mathcal L\!\left(
       \left.\frac{d}{dz}\Gamma_\sigma^{-z}(A^z)\right|_{z=1}
       \right)\right]}_{=0}\\
 &=\Tr\!\left[(\mathcal L^\sharp)_*(A)
                    (\log A-\log\sigma)\right].
 \end{aligned}
\]
Multiplying by $-1/2$ and applying Proposition~\ref{prop:endpoint}
proves \eqref{eq:generator-endpoint}.
\end{proof}

\section{\texorpdfstring{$L_p$}{Lp}-regularity bounds for channels and semigroups}
\label{sec:comparison-bounds}

We first establish a finite-channel bound in Cayley form by Poisson
representation on a strip. Its small-time limit gives $L_p$-regularity
bounds for QMS generators. The same first variation gives the properties of
$D_A^{\mathcal L}$
used in Section~\ref{sec:sharpness}.

\subsection{Finite-channel bounds in Cayley form}
\label{sec:channels}

For a UCP map $\Phi$ preserving $\sigma$ and a faithful state $\rho$,
we study the channel function $F_{\Phi,\rho}$ from
\eqref{eq:channel-function} on the closed strip $0\le\Ree z\le1$.
Its Cayley transform is
\[
 G_{\Phi,\rho}(z)=\frac{1-F_{\Phi,\rho}(z)}{1+F_{\Phi,\rho}(z)}
\]
whenever $1+F_{\Phi,\rho}(z)\ne0$.
\begin{theorem}
\label{thm:channel}
Let $\Phi$ be a UCP map with faithful invariant state $\sigma$ and
$\rho$ be a faithful state. For every $0<\alpha<1$,
$0\le F_{\Phi,\rho}(\alpha)\le1$ and
\begin{equation}
 \frac{1-F_{\Phi,\rho}(\alpha)}{1+F_{\Phi,\rho}(\alpha)}
 \ge \theta_{\mathrm G}(\alpha)
       \frac{1-F_{\Phi,\rho}(1/2)}{1+F_{\Phi,\rho}(1/2)},
 \label{eq:channel-cayley}
\end{equation}
where
\begin{equation}
 \theta_{\mathrm G}(\alpha)
 =\tan\!\left(\frac\pi2\min\{\alpha,1-\alpha\}\right).
 \label{eq:channel-coefficients}
\end{equation}
If $\Phi=\Phi^\sharp$, the same inequality holds with
$ \theta_{\mathrm K}(\alpha)=\sin(\pi\alpha)$ in place of $\theta_{\mathrm G}$.
\end{theorem}

We first derive the strip bounds and the identity
\eqref{eq:channel-reflection} from the channel hypotheses.
In the proof of Theorem~\ref{thm:channel}, we apply the Poisson formulas
\eqref{eq:strip-poisson} and~\eqref{eq:poisson-midpoint} to bounded
regularizations of the Cayley transform and pass to
the limit to obtain \eqref{eq:channel-cayley}. In the KMS case, the symmetry
$G_{\Phi,\rho}(z)=G_{\Phi,\rho}(1-z)$ yields the improved sine constant.
\begin{lemma}
\label{lem:strip}
$F_{\Phi,\rho}(z)$, defined in \eqref{eq:channel-function},
is continuous on
the closed strip $0\le\Ree z\le1$, analytic in its interior, and satisfies
\[
 |F_{\Phi,\rho}(z)|\le1\quad(0\le\Ree z\le1).
\]
It is nonnegative on $[0,1]$, with
$F_{\Phi,\rho}(0)=F_{\Phi,\rho}(1)=1$. Moreover,
\begin{equation}
 F_{\Phi,\rho}(z)=F_{\Phi^\sharp,\rho}(1-z).
 \label{eq:channel-reflection}
\end{equation}
\end{lemma}

\begin{proof}
Using $\Phi^\sharp=\Gamma_\sigma^{-1}\Phi_*\Gamma_\sigma$ and cyclicity, we obtain
\[
 \begin{split}
 F_{\Phi^\sharp,\rho}(1-z)
 &=\Tr\left[\Gamma_\sigma^{-z}(\rho^z)
             \Phi_*\Gamma_\sigma^z(\rho^{1-z})\right]\\
 &=\Tr\left[\Gamma_\sigma^z(\rho^{1-z})
             \Phi\Gamma_\sigma^{-z}(\rho^z)\right]\\
 &=\Tr\left[\rho^{1-z}\Gamma_\sigma^z
             \Phi\Gamma_\sigma^{-z}(\rho^z)\right]\\
 &=F_{\Phi,\rho}(z).
 \end{split}
\]

As noted after \eqref{eq:analytic-energy}, $F_{\Phi,\rho}$ is entire
and bounded on the closed strip.
For $z=iy$, the matrix $\Gamma_\sigma^{-iy}(\rho^{iy})$ is unitary.
Operator-norm contraction of $\Phi$ and trace H\"older therefore give
\[
 |F_{\Phi,\rho}(iy)|
 \le \|\rho^{1-iy}\|_1
      \|\Phi\Gamma_\sigma^{-iy}(\rho^{iy})\|_\infty
 \le1.
\]
The map $\Phi^\sharp$ is also UCP and preserves
$\sigma$. Applying the same estimate to $\Phi^\sharp$ and using
\eqref{eq:channel-reflection} yields
\[
 |F_{\Phi,\rho}(1+iy)|=|F_{\Phi^\sharp,\rho}(-iy)|\le1.
\]
Since $F_{\Phi,\rho}$ is bounded on the closed strip, the Hadamard
three-lines theorem \cite[Theorem~12.2.1]{Wilde17} and the two boundary
estimates give $|F_{\Phi,\rho}(z)|\le1$ throughout the strip.

For real $\alpha$, \eqref{eq:channel-function} is the trace of a product
of two positive matrices, so it is nonnegative. Unitality gives
$F_{\Phi,\rho}(0)=1$, and \eqref{eq:channel-reflection} gives
$F_{\Phi,\rho}(1)=F_{\Phi^\sharp,\rho}(0)=1$.
\end{proof}

To compare the Cayley transform with its midpoint value, we first
record the Poisson formula on the strip.
\begin{lemma}\label{lem:strip-poisson}
Let $H$ be bounded and continuous on $0\le\Ree z\le1$ and analytic
in the interior. Write $u_b(t)=\Ree H(b+it)$ for $b=0,1$. For $0<\alpha<1$,
\begin{equation}\label{eq:strip-poisson}
 \Ree H(\alpha)=\frac{\sin(\pi \alpha)}2\int_{\mathbb R}
 \left(\frac{u_0(t)}{\cosh(\pi t)-\cos(\pi \alpha)}
       +\frac{u_1(t)}{\cosh(\pi t)+\cos(\pi \alpha)}\right)dt.
\end{equation}
In particular,
\begin{equation}\label{eq:poisson-midpoint}
 \Ree H(1/2)=\frac12\int_{\mathbb R}
                       \frac{u_0(t)+u_1(t)}{\cosh(\pi t)}dt.
\end{equation}
\end{lemma}
\begin{proof}
For a bounded harmonic function $v$ on the upper half-plane that
extends continuously to $\mathbb R\setminus\{0\}$,
the Poisson representation
\cite[Theorems~7.17(b) and~7.28]{ABR01} gives
\[
 v(x+iy)=\frac1\pi\int_{\mathbb R}
 \frac{y\,v(\lambda)}{(x-\lambda)^2+y^2}\,d\lambda,
 \qquad y>0.
\]
Here $v(\lambda)$ denotes the continuous boundary value for
$\lambda\ne0$; its value at $0$ may be chosen arbitrarily.

Set
\[
 U(w)=\Ree H\!\left(\frac{\operatorname{Log}w}{i\pi}\right),
 \qquad \Im w>0.
\]
Here $\operatorname{Log}w=\log|w|+i\arg w$, with $0<\arg w<\pi$.
The map $w\mapsto\operatorname{Log}w/(i\pi)$ sends the upper half-plane
conformally onto the strip $0<\Ree z<1$.
The function $w\mapsto H(\operatorname{Log}w/(i\pi))$ is holomorphic on
the upper half-plane, so its real part $U$ is harmonic. Together with
boundedness of $H$, this gives
\[
 \Delta U=0,\qquad
 \sup_{\Im w>0}|U(w)|
 \le\sup_{0\le\Ree z\le1}|H(z)|<\infty.
\]
By continuity of $H$ on the closed strip, $U$ extends continuously to
$\mathbb R\setminus\{0\}$, with boundary values
\[
 U\!\left((-1)^b e^{-\pi t}\right)=u_b(t),
 \qquad b=0,1,\quad t\in\mathbb R.
\]
For $0<\alpha<1$, this gives the Poisson representation for $\Ree H$:
\[
 \begin{aligned}
 \Ree H(\alpha)
 &=U(e^{i\pi \alpha})=\frac{\sin(\pi \alpha)}\pi
   \int_{\mathbb R}\frac{U(\lambda)}{|e^{i\pi \alpha}-\lambda|^2}\,d\lambda\\
 &\overset{\lambda=(-1)^b e^{-\pi t}}{=}
   \frac{\sin(\pi \alpha)}\pi\sum_{b=0}^1\int_{\mathbb R}
   \frac{\pi e^{-\pi t}u_b(t)}
        {|e^{i\pi \alpha}-(-1)^b e^{-\pi t}|^2}\,dt\\
 &=\frac{\sin(\pi \alpha)}2\int_{\mathbb R}
   \left(\frac{u_0(t)}{\cosh(\pi t)-\cos(\pi \alpha)}
        +\frac{u_1(t)}{\cosh(\pi t)+\cos(\pi \alpha)}\right)dt.
 \end{aligned}
\]
The last equality follows from
\[
 \begin{aligned}
 |e^{i\pi \alpha}-(-1)^b e^{-\pi t}|^2
 &=1+e^{-2\pi t}-2(-1)^b e^{-\pi t}\cos(\pi \alpha)\\
 &=2e^{-\pi t}\bigl[\cosh(\pi t)-(-1)^b\cos(\pi \alpha)\bigr].
 \end{aligned}
\]
This proves \eqref{eq:strip-poisson}. Taking $\alpha=1/2$ gives
\eqref{eq:poisson-midpoint}.
\end{proof}

We now prove Theorem~\ref{thm:channel} by applying the Poisson formulas
of Lemma~\ref{lem:strip-poisson} to a regularized Cayley transform of
$F_{\Phi,\rho}$.

\begin{proof}[Proof of Theorem~\ref{thm:channel}]
For $0<r<1$, put
\[
 G_r(z)=\frac{1-rF_{\Phi,\rho}(z)}{1+rF_{\Phi,\rho}(z)}.
\]
By Lemma~\ref{lem:strip}, $G_r$ is continuous on the closed strip,
analytic in its interior, and satisfies
\[
 |G_r(z)|\le\frac{1+r}{1-r},\qquad
 \Ree G_r(z)=\frac{1-r^2|F_{\Phi,\rho}(z)|^2}{|1+rF_{\Phi,\rho}(z)|^2}>0.
\]
It is positive on $(0,1)$. Write $u_b(t)=\Ree G_r(b+it)$ for $b=0,1$.
For $0<\alpha<1$,
\eqref{eq:strip-poisson} and \eqref{eq:poisson-midpoint}, applied
to $H=G_r$, give
\[
 \begin{aligned}
 G_r(\alpha)
 &=\frac{\sin(\pi \alpha)}2\int_{\mathbb R}
 \left(\frac{u_0(t)}{\cosh(\pi t)-\cos(\pi \alpha)}
       +\frac{u_1(t)}{\cosh(\pi t)+\cos(\pi \alpha)}\right)dt\\
 &\ge\frac{\sin(\pi \alpha)}{2(1+|\cos(\pi \alpha)|)}
 \int_{\mathbb R}
 \frac{u_0(t)+u_1(t)}{\cosh(\pi t)}\,dt\\
 &=\frac{\sin(\pi \alpha)}{1+|\cos(\pi \alpha)|}G_r(1/2)
 =\theta_{\mathrm G}(\alpha)G_r(1/2).
 \end{aligned}
\]
The inequality uses the bound
\[
 0<\cosh(\pi t)-(-1)^b\cos(\pi \alpha)
 \le(1+|\cos(\pi \alpha)|)\cosh(\pi t),\qquad b=0,1.
\]
If $\Phi=\Phi^\sharp$, \eqref{eq:channel-reflection} gives
$G_r(z)=G_r(1-z)$. Since $G_r$ is real on $(0,1)$, the identity
theorem and continuity give $G_r(\bar z)=\overline{G_r(z)}$ on the
closed strip. Hence
\[
 u_1(t)=\Ree G_r(1+it)=\Ree\overline{G_r(it)}=u_0(t)\ge0.
\]
Equations~\eqref{eq:strip-poisson} and~\eqref{eq:poisson-midpoint} then give
\[
 \begin{aligned}
 G_r(\alpha)
 &=\sin(\pi \alpha)\int_{\mathbb R}
   \frac{\cosh(\pi t)u_0(t)}
        {\cosh^2(\pi t)-\cos^2(\pi \alpha)}\,dt\\
 &\ge\sin(\pi \alpha)\int_{\mathbb R}
   \frac{u_0(t)}{\cosh(\pi t)}\,dt
 =\sin(\pi \alpha)G_r(1/2).
 \end{aligned}
\]
The inequality follows from
$0<\cosh^2(\pi t)-\cos^2(\pi \alpha)\le\cosh^2(\pi t)$.
Since $1+F_{\Phi,\rho}(\alpha)\ge1$ for $0<\alpha<1$, letting $r\to1^-$ in these
inequalities proves \eqref{eq:channel-cayley} and its KMS improvement,
including the case $F_{\Phi,\rho}\equiv1$.

\end{proof}

\begin{remark}\label{rem:channel-energy}
For a UCP map $\Phi$ with faithful invariant state $\sigma$, let
$\mathcal E_{p,\Phi}=\mathcal E_p^{\Phi-\id}$ denote the $L_p$ Dirichlet forms
of the QMS generator $\Phi-\id$.
Theorem~\ref{thm:channel} yields, for $1<p<\infty$ and $f>0$,
\begin{equation}
 \mathcal E_{p,\Phi}(f)
 \ge
 \frac{\kappa_{\mathrm G}(p)\mathcal E_{2,\Phi}(I_{2,p}f)}
 {1-\dfrac{1-\theta_{\mathrm G}(1/p)}{2\|f\|_{p,\sigma}^p}
          \mathcal E_{2,\Phi}(I_{2,p}f)}.
 \label{eq:channel-energy}
\end{equation}
If $\Phi=\Phi^\sharp$, the same estimate holds with
$\kappa_{\mathrm K},\theta_{\mathrm K}$ in place of
$\kappa_{\mathrm G},\theta_{\mathrm G}$.
At $p=2$ the estimate is an identity.
The estimate follows by applying Theorem~\ref{thm:channel} to $A/(\Tr A)$,
where $A=(\Gamma_\sigma^{1/p}f)^p$, and using
$D_A^{\Phi-\id}(z)=\Tr A\,[1-F_{\Phi,A/(\Tr A)}(z)]$ and \eqref{eq:energy-transport}.
\end{remark}

\subsection{\texorpdfstring{$L_p$}{Lp}-regularity for generators}
\label{sec:comparisons}

A small-time limit of Theorem~\ref{thm:channel} gives the
$L_p$-regularity bounds for QMS generators. The next lemma identifies
$D_\rho^{\mathcal L}$ as the negative first variation of the
channel function $F_{T_t,\rho}$ at $t=0$, and records the strip
properties used in Section~\ref{sec:sharpness}.

\begin{lemma}
\label{lem:generator-transport}
Let $T_t=e^{t\mathcal L}$ be a QMS with faithful
invariant state $\sigma$. For a faithful state $\rho$ and
$z$ in the closed strip $0\le\Ree z\le1$,
\begin{equation}
 F_{T_t,\rho}(z)=1-tD_\rho^{\mathcal L}(z)+O(t^2),\qquad
 G_{T_t,\rho}(z)=\frac t2D_\rho^{\mathcal L}(z)+O(t^2)
 \quad(t\to0^+).
 \label{eq:small-time-expansion}
\end{equation}
For every $A>0$, the function
$D_A^{\mathcal L}$ is analytic and bounded on the closed strip,
real on $(0,1)$, and satisfies
\[
 \Ree D_A^{\mathcal L}(z)\ge0,\qquad
 D_A^{\mathcal L}(0)=D_A^{\mathcal L}(1)=0.
\]
\end{lemma}
\begin{proof}
Fix a faithful state $\rho$ and $z$ with $0\le\Ree z\le1$.
The exponential series gives
$T_t=\id+t\mathcal L+O(t^2)$. For these fixed
$\rho,z$, the map $M\mapsto F_{M,\rho}(z)$ is a bounded linear
functional. Thus \eqref{eq:channel-function} and
\eqref{eq:analytic-energy} give
\[
 \begin{aligned}
 F_{T_t,\rho}(z)
 &=F_{\id,\rho}(z)+tF_{\mathcal L,\rho}(z)+O(t^2)\\
 &=\Tr\rho-tD_\rho^{\mathcal L}(z)+O(t^2)
  =1-tD_\rho^{\mathcal L}(z)+O(t^2).
 \end{aligned}
\]
The Cayley transform therefore satisfies
\[
 G_{T_t,\rho}(z)
 =\frac{tD_\rho^{\mathcal L}(z)+O(t^2)}
        {2-tD_\rho^{\mathcal L}(z)+O(t^2)}
 =\frac t2D_\rho^{\mathcal L}(z)+O(t^2).
\]
This proves \eqref{eq:small-time-expansion}.

For each $t>0$, Lemma~\ref{lem:strip}, applied to the channel $T_t$,
gives $\Ree F_{T_t,\rho}(z)\le|F_{T_t,\rho}(z)|\le1$.
Taking real parts in the first expansion in
\eqref{eq:small-time-expansion} yields
\[
 \Ree D_\rho^{\mathcal L}(z)
 =\lim_{t\to0^+}\frac{1-\Ree F_{T_t,\rho}(z)}t\ge0.
\]
By homogeneity, $\Ree D_A^{\mathcal L}(z)\ge0$ for every $A>0$.

The analyticity and boundedness of $D_A^{\mathcal L}$ were noted
after \eqref{eq:analytic-energy}.
Lemma~\ref{lem:strip} also gives real values of $F_{T_t,\rho}$
on $(0,1)$ and $F_{T_t,\rho}(0)=F_{T_t,\rho}(1)=1$.
The first expansion in \eqref{eq:small-time-expansion} therefore
shows that $D_\rho^{\mathcal L}$ is real on $(0,1)$ and vanishes
at $0,1$. Homogeneity gives these properties for every $A>0$.
\end{proof}

\begin{theorem}
\label{thm:comparison}
Let $\mathcal H$ be a finite-dimensional Hilbert space and $T_t=e^{t\mathcal L}$ be a
QMS on $B(\mathcal H)$ with faithful invariant state $\sigma$.
For every $1\le p<\infty$ and $f>0$,
\begin{equation}
 \mathcal E_p^{\mathcal L}(f)
 \ge\kappa_{\mathrm G}(p)
       \mathcal E_2^{\mathcal L}(I_{2,p}f),
 \label{eq:general-comparison}
\end{equation}
where
\[
 \kappa_{\mathrm G}(p)
 =\frac{p}{2(p-1)}
 \tan\!\left(\frac{\pi}{2}\min\{1/p,1-1/p\}\right),
 \qquad 1<p<\infty,
\]
and $\kappa_{\mathrm G}(1)=\pi/4$.
If $\mathcal L$ is KMS symmetric, then
\begin{equation}
 \mathcal E_p^{\mathcal L}(f)
 \ge\kappa_{\mathrm K}(p)
       \mathcal E_2^{\mathcal L}(I_{2,p}f),
 \label{eq:kms-comparison}
\end{equation}
where
\[
 \kappa_{\mathrm K}(p)
 =\frac{p\sin(\pi/p)}{2(p-1)},
 \qquad 1<p<\infty,
\]
and $\kappa_{\mathrm K}(1)=\pi/2$.
\end{theorem}
\begin{proof}[Proof of Theorem~\ref{thm:comparison}]
Fix $1<p<\infty$ and $f>0$. Put
$A=(\Gamma_\sigma^{1/p}f)^p$ and $\rho=A/(\Tr A)$.
Applying Theorem~\ref{thm:channel} to $\Phi=T_t$ yields:
\[
 G_{T_t,\rho}(1/p)\ge\theta_{\mathrm G}(1/p)G_{T_t,\rho}(1/2).
\]
Divide by $t/2$ and let $t\to0^+$. Equation~\eqref{eq:small-time-expansion}
and homogeneity give
\[
 D_A^{\mathcal L}(1/p)\ge
 \theta_{\mathrm G}(1/p)D_A^{\mathcal L}(1/2).
\]
By \eqref{eq:energy-transport}, this proves \eqref{eq:general-comparison},
since $\kappa_{\mathrm G}(p)=p\theta_{\mathrm G}(1/p)/[2(p-1)]$.

If $\mathcal L$ is KMS symmetric, then so is $T_t$, and
Theorem~\ref{thm:channel} gives
\[
 G_{T_t,\rho}(1/p)\ge\theta_{\mathrm K}(1/p)G_{T_t,\rho}(1/2).
\]
The same small-time limit and \eqref{eq:energy-transport} give
\eqref{eq:kms-comparison}, since
$\kappa_{\mathrm K}(p)=p\theta_{\mathrm K}(1/p)/[2(p-1)]$.

The case $p=1$ follows by letting $p\to1^+$.
\end{proof}

\begin{remark}
\label{rem:physical-endpoint}
The $p=1$ endpoint of Theorem~\ref{thm:comparison} recovers the
entropy-production bounds of \cite[Theorem~5.1]{GW26conditional}.
For $A>0$, set $f=\Gamma_\sigma^{-1}A$ and
$g=I_{2,1}f=\Gamma_\sigma^{-1/2}\sqrt A$.
By Propositions~\ref{prop:endpoint} and~\ref{prop:duality},
$\mathcal I_{\mathcal L,\sigma}(A)=2\mathcal E_1^{\mathcal L^\sharp}(f)$
and $\mathcal E_2^{\mathcal L^\sharp}(g)=\mathcal E_2^{\mathcal L}(g)$.
Applying the endpoint comparison to $\mathcal L^\sharp$ gives
\[
 \mathcal I_{\mathcal L,\sigma}(A)
 \ge\frac\pi2\mathcal E_2^{\mathcal L}(g),
\]
with constant $\pi$ under KMS symmetry.
Here $\mathcal E_2^{\mathcal L}(g)$ is the square-root energy used
in the cited theorem; the extension from faithful states to $A>0$
follows by homogeneity.
\end{remark}

\section{Sharpness of the \texorpdfstring{$L_p$}{Lp}-regularity bounds}
\label{sec:sharpness}

Theorem~\ref{thm:comparison} gives universal lower bounds for the
energy quotient. We now construct matching tests in the general and
KMS classes, proving the optimality of the constants in
Theorem~\ref{thm:optimal-class}.

\begin{theorem}\label{thm:sharpness}
For every $1\le p<\infty$,
\begin{equation}\label{eq:optimal-coefficient}
 \begin{aligned}
 \inf_{\mathcal L,\sigma,f}
 \frac{\mathcal E_p^{\mathcal L}(f)}
      {\mathcal E_2^{\mathcal L}(I_{2,p}f)}
 &=\kappa_{\mathrm G}(p),\\
 \inf_{\substack{\mathcal L,\sigma,f\\
                  \mathcal L=\mathcal L^\sharp}}
 \frac{\mathcal E_p^{\mathcal L}(f)}
      {\mathcal E_2^{\mathcal L}(I_{2,p}f)}
 &=\kappa_{\mathrm K}(p),
 \end{aligned}
\end{equation}
where both infima range over all nonzero finite-dimensional Hilbert
spaces $\mathcal H$, QMS generators $\mathcal L$ on $B(\mathcal H)$,
their faithful invariant states $\sigma$, and $f>0$ with
$\mathcal E_2^{\mathcal L}(I_{2,p}f)>0$.
The constants $\kappa_{\mathrm G}$ and $\kappa_{\mathrm K}$ are defined
in Theorem~\ref{thm:comparison}.
\end{theorem}

\begin{proof}
It remains to establish the sharpness of the constants in
Theorem~\ref{thm:comparison}.
We first construct finite-dimensional generators and rank-one inputs.
The case $p=1$ requires a separate calculation of entropy production.

\emph{A common model and its boundary values.}
Fix $h>0$, an integer $n\ge2$, and $\epsilon=-1$ or $1$ for the general or KMS case.
Define
\[
 B=\operatorname{diag}(\lambda_j),\qquad
 \lambda_j=\epsilon^je^{2jh},\quad 0\le j\le n,
 \qquad |B|=\operatorname{diag}(e^{2jh}).
\]
Let $\mathbf1=(1,\ldots,1)^T\in\mathbb C^{n+1}$ and set
\[
 V=\frac{(B\mathbf1)\mathbf1^T}{\|B\|_2},
 \qquad K_{ij}=\frac{\lambda_j}{\lambda_i+\lambda_j}.
\]
With $F(u,v)=(v,u)$ on $\mathbb C^2$, set
\[
 W=V\otimes F,\qquad K_0=K\otimes I_2,\qquad
 H=\frac i2(K_0-K_0^*).
\]
Define the generator on $B(\mathbb C^{n+1}\otimes\mathbb C^2)$
in GKSL form by
\[
 \mathcal L(X)=i[H,X]+W^*XW
 -\frac12\bigl(W^*WX+XW^*W\bigr).
\]
A faithful invariant state of the resulting QMS is
\[
 \sigma=\frac{B^2\otimes I_2}{2\Tr B^2}.
\]
Indeed,
\begin{align}
 K+K^*&=\mathbf1\mathbf1^T=V^*V,
 \label{eq:model-unital}\\
 K^*B^2+B^2K&=(B\mathbf1)(B\mathbf1)^T=VB^2V^*.
 \label{eq:model-invariant}
\end{align}
Equation~\eqref{eq:model-unital} gives $K_0+K_0^*=W^*W$, so the
generator simplifies to
\[
 \mathcal L(X)=W^*XW-K_0X-XK_0^*.
\]
Equation~\eqref{eq:model-invariant} gives
\[
 \mathcal L_*(\sigma)
 =\frac{(VB^2V^*-K^*B^2-B^2K)\otimes I_2}{2\Tr B^2}
 =0.
\]
Since $B$ is invertible, $\sigma>0$; its normalization gives
$\Tr\sigma=1$. Moreover, $B^2=\operatorname{diag}(e^{4jh})$, so
$\sigma$ is the same for both choices of $\epsilon$.
For $\epsilon=1$, we have $B>0$ and
\[
 \begin{aligned}
 \mathcal L^\sharp(X)
 &=\sigma^{-1/2}\mathcal L_*(\sigma^{1/2}X\sigma^{1/2})\sigma^{-1/2}\\
 &=\bigl[(BV^*B^{-1})\otimes F\bigr]^*
   X\bigl[(BV^*B^{-1})\otimes F\bigr]\\
 &\quad-\bigl[(B^{-1}K^*B)\otimes I_2\bigr]X
       -X\bigl[(B^{-1}K^*B)\otimes I_2\bigr]^*.
 \end{aligned}
\]
The definitions of $V$ and $K$ give
\[
 BV^*B^{-1}=V,\qquad B^{-1}K^*B=K.
\]
Substituting these identities into the expansion above gives
$\mathcal L^\sharp(X)=\mathcal L(X)$ for every $X$, so
$\mathcal L^\sharp=\mathcal L$.

Take the real unit vector and rank-one projection
\[
 \psi_j=\frac{(-1)^j\binom nj}
                  {\bigl(\sum_{k=0}^n\binom nk^2\bigr)^{1/2}},
 \qquad \widehat\psi=\psi\otimes e_0,\quad e_0=(1,0)^T,
 \qquad P=|\widehat\psi\rangle\langle\widehat\psi|.
\]
The binomial theorem gives $\sum_j\psi_j=0$, so
\[
 V\psi=\frac{B\mathbf1}{\|B\|_2}(\mathbf1^T\psi)=0,
 \qquad W\widehat\psi=(V\psi)\otimes(Fe_0)=0.
\]
Define
\[
 \begin{aligned}
 d(z)&=-\Tr\!\left[\sigma^{z/2}P\sigma^{z/2}
              \mathcal L(\sigma^{-z/2}P\sigma^{-z/2})\right]\\
     &=2\langle\psi,|B|^zK|B|^{-z}\psi\rangle.
 \end{aligned}
\]
Thus $d$ is entire, bounded on
the closed strip, and real on the real axis.
Since $K$ and $\psi$ are real, \eqref{eq:model-unital} gives
\[
 d(0)=2\psi^TK\psi=\psi^T(K+K^*)\psi
     =|\mathbf1^T\psi|^2=0.
\]

Let $u_b(t)=\Ree d(b+it)$ for $b=0,1$ be the real parts
of $d$ on the boundary lines $\Ree z=b$.
Equations~\eqref{eq:model-unital} and~\eqref{eq:model-invariant} give
\[
 \begin{aligned}
 u_0(t)&=\left|\sum_{j=0}^n\psi_j e^{2ijht}\right|^2,\\
 u_1(t)&=\left|\sum_{j=0}^n\epsilon^j\psi_j e^{2ijht}\right|^2.
 \end{aligned}
\]
Indeed, the left boundary uses $K+K^*=\mathbf1\mathbf1^T$;
on the right boundary the relevant matrix is
$|B|^{-1}(K^*B^2+B^2K)|B|^{-1}$, whose $(i,j)$ entry is
$\epsilon^i\epsilon^j$.
Now
\[
 \sum_{j=0}^n(-1)^j\binom nj e^{2ijht}=(1-e^{2iht})^n,
 \qquad
 \sum_{j=0}^n\binom nj e^{2ijht}=(1+e^{2iht})^n.
\]
Consequently, with
\[
 a_n(t)=\frac{4^n\sin^{2n}(ht)}{\sum_{j=0}^n\binom nj^2},\qquad
 b_n(t)=\frac{4^n\cos^{2n}(ht)}{\sum_{j=0}^n\binom nj^2},
\]
the boundary values are
\[
 u_0(t)=a_n(t),\qquad
 u_1(t)=
 \begin{cases}
 b_n(t),&\epsilon=-1,\\
 a_n(t),&\epsilon=1.
 \end{cases}
\]
Thus the choice of $\epsilon$ changes the real part $u_1$ on the
right boundary $\Ree z=1$, while leaving $u_0$ unchanged.

\emph{Periodic sampling and the scalar ratios.}
First fix $h$ and let $n\to\infty$. Put
$T=\frac\pi h$. Orthogonality of the exponentials and
$\sum_j|\psi_j|^2=1$ give
\[
 \int_{-T/2}^{T/2}a_n(t)\,dt
 =\int_{-T/2}^{T/2}b_n(t)\,dt=T.
\]
For every continuous exponentially decaying function $G$,
\begin{align}
 \lim_{n\to\infty}\frac1T\int_{\mathbb R}b_n(t)G(t)\,dt
     &=\sum_{k\in\mathbb Z}G(kT),\label{eq:sampling-cosine}\\
 \lim_{n\to\infty}\frac1T\int_{\mathbb R}a_n(t)G(t)\,dt
     &=\sum_{k\in\mathbb Z}G((k+1/2)T).\label{eq:sampling-sine}
\end{align}
To see this, periodize $G$:
\[
 G_T(t)=\sum_{k\in\mathbb Z}G(t+kT),\qquad
 \int_{\mathbb R}b_nG
   =\int_{-T/2}^{T/2}b_nG_T.
\]
Exponential decay makes the series uniformly convergent on
$[-T/2,T/2]$. The normalized cosine powers concentrate at zero:
for $0<r<T/2$, the periodic normalization gives
\[
 \begin{aligned}
 \frac1T\int_{\{r\le|t|\le T/2\}}b_n(t)\,dt
 &=\frac{\displaystyle\int_{\{r\le|t|\le T/2\}}
                    \cos^{2n}(ht)\,dt}
         {\displaystyle\int_{-T/2}^{T/2}\cos^{2n}(ht)\,dt}\\
 &\le\frac{T\cos^{2n}(hr)}
            {\displaystyle\int_{-r/2}^{r/2}\cos^{2n}(ht)\,dt}\\
 &\le\frac{T\cos^{2n}(hr)}{r\cos^{2n}(hr/2)}
 =\frac Tr\left[\frac{\cos(hr)}{\cos(hr/2)}\right]^{2n}.
 \end{aligned}
\]
Since $0<\cos(hr)/\cos(hr/2)<1$, the right-hand side tends to zero as
$n\to\infty$, so
\[
 \lim_{n\to\infty}\frac1T
 \int_{\{r\le|t|\le T/2\}}b_n(t)\,dt=0.
\]
Integration against the continuous $G_T$ therefore gives $G_T(0)$.
Translation by half a period proves the sine formula
\eqref{eq:sampling-sine}.

Set
\[
 G(t):=\frac1{\cosh(\pi t)-1},
 \qquad t\in\mathbb R\setminus\{0\}.
\]
This function has the exponential decay at infinity required by
\eqref{eq:sampling-sine}, since $G(t)=O(e^{-\pi|t|})$ as
$|t|\to\infty$; however, it is singular at
zero. To establish the corresponding sampling limit
\begin{equation}\label{eq:sampling-singular}
 \lim_{n\to\infty}\frac1T
 \int_{\mathbb R}\frac{a_n(t)}{\cosh(\pi t)-1}\,dt
 =\sum_{k\in\mathbb Z}
   \frac1{\cosh(\pi(k+1/2)T)-1},
\end{equation}
define
\[
 \widetilde G_h(t)=\frac{\sin^2(ht)}{\cosh(\pi t)-1}
 \quad(t\ne0),\qquad
 \text{where }\widetilde G_h(0)=\frac{2h^2}{\pi^2}.
\]
Thus $\widetilde G_h$ is continuous. The bound $\sin^2(ht)\le1$
also gives exponential decay. We have
\[
 \frac1T\int_{\mathbb R}a_n(t)G(t)\,dt
 =\frac{2n}{2n-1}\,
   \frac1T\int_{\mathbb R}a_{n-1}(t)\widetilde G_h(t)\,dt.
\]
Indeed, the definition of $a_n$ gives
\[
 \begin{aligned}
 a_n(t)&=\frac{4^n\sin^{2n}(ht)}{\binom{2n}{n}}\\
       &=\frac{4\binom{2n-2}{n-1}}{\binom{2n}{n}}
         \sin^2(ht)\,a_{n-1}(t)\\
       &=\frac{4n^2}{(2n)(2n-1)}\sin^2(ht)\,a_{n-1}(t)\\
       &=\frac{2n}{2n-1}\sin^2(ht)\,a_{n-1}(t).
 \end{aligned}
\]
The first two equalities use
$\sum_{k=0}^j\binom jk^2=\binom{2j}{j}$ for $j=n,n-1$,
which follows by comparing $x^j$ coefficients in
$(1+x)^j(1+x)^j=(1+x)^{2j}$.
Also, $\widetilde G_h((k+1/2)T)=G((k+1/2)T)$, since
$\sin^2(h(k+1/2)T)=1$. Applying \eqref{eq:sampling-sine} to
$\widetilde G_h$ and using
$\lim_{n\to\infty}\frac{2n}{2n-1}=1$ therefore proves
\eqref{eq:sampling-singular}.

Fix $0<\alpha<1$ and put $C(t)=\cosh(\pi t)$.
We use the following lattice estimates as $h\to0^+$, so that
$T\to\infty$. For each fixed $\gamma\in[-1,1)$,
\begin{align}
 \lim_{h\to0^+}\sum_{k\in\mathbb Z}\frac1{C(kT)-\gamma}
       &=\frac1{1-\gamma},\label{eq:lattice-integer-limit}\\
 \sum_{k\in\mathbb Z}\frac1{C((k+1/2)T)-\gamma}
       &=O(e^{-\pi T/2}).\label{eq:lattice-half-integer-tail}
\end{align}
The integer limit \eqref{eq:lattice-integer-limit} follows by separating
the $k=0$ term and summing the exponentially decaying tails.
The half-integer estimate \eqref{eq:lattice-half-integer-tail} also holds
for $\gamma=1$, since these points stay at distance at least $T/2$ from zero.

For the general case, apply Lemma~\ref{lem:strip-poisson}
with $H=d$, $u_0=a_n$, and $u_1=b_n$.
For $0<\alpha<1$, \eqref{eq:strip-poisson} gives
\[
 d(\alpha)=\frac{\sin(\pi\alpha)}2\int_{\mathbb R}
   \left(\frac{a_n}{C-\cos(\pi\alpha)}
        +\frac{b_n}{C+\cos(\pi\alpha)}\right)dt.
\]
Apply \eqref{eq:sampling-sine} to the $a_n$ integral with
$G(t)=1/[C(t)-\cos(\pi\alpha)]$, and
\eqref{eq:sampling-cosine} to the $b_n$ integral with
$G(t)=1/[C(t)+\cos(\pi\alpha)]$. Dividing the interior identity by $T$ gives
\[
 \lim_{n\to\infty}\frac{d(\alpha)}T
 =\frac{\sin(\pi\alpha)}2\sum_{k\in\mathbb Z}
   \left(\frac1{C((k+1/2)T)-\cos(\pi\alpha)}
        +\frac1{C(kT)+\cos(\pi\alpha)}\right).
\]
In \eqref{eq:lattice-half-integer-tail} and
\eqref{eq:lattice-integer-limit}, take
$\gamma=\cos(\pi\alpha)$ and $\gamma=-\cos(\pi\alpha)$,
respectively. Then
\[
 \lim_{h\to0^+}\lim_{n\to\infty}\frac{d(\alpha)}T
 =\frac{\sin(\pi\alpha)}2
   \left(0+\frac1{1+\cos(\pi\alpha)}\right)
 =\frac{\sin(\pi\alpha)}{2(1+\cos(\pi\alpha))}.
\]
Taking $\alpha=1/2$ gives the midpoint limit
\[
 \lim_{h\to0^+}\lim_{n\to\infty}\frac{d(1/2)}T
 =\frac1{2(1+0)}=\frac12.
\]

For the endpoint derivative, apply \eqref{eq:strip-poisson} of
Lemma~\ref{lem:strip-poisson} with interior parameter $0<\beta<1$.
In each fixed finite model,
$d(0)=0$ and $a_n(t)=O(t^2)$ near zero, so dominated convergence gives
\begin{equation}\label{eq:poisson-endpoint-right}
 \begin{aligned}
 \frac{d'(0)}2
 &=\lim_{\beta\to0^+}\frac{d(\beta)}{2\beta}\\
 &=\lim_{\beta\to0^+}\frac{\sin(\pi\beta)}{4\beta}
   \int_{\mathbb R}
   \left(\frac{a_n}{C-\cos(\pi\beta)}
        +\frac{b_n}{C+\cos(\pi\beta)}\right)dt\\
 &=\frac\pi4\int_{\mathbb R}
   \left(\frac{a_n}{C-1}+\frac{b_n}{C+1}\right)dt.
 \end{aligned}
\end{equation}
For the $a_n$ integral, use \eqref{eq:sampling-singular} with the
singular weight $G(t)=1/(C(t)-1)$. For the $b_n$ integral, use
\eqref{eq:sampling-cosine} with $G(t)=1/(C(t)+1)$.
Dividing the endpoint identity by $T$ and taking $n\to\infty$
at fixed $h$ gives
\[
 \lim_{n\to\infty}\frac{d'(0)}{2T}
 =\frac\pi4\sum_{k\in\mathbb Z}
   \left(\frac1{C((k+1/2)T)-1}+\frac1{C(kT)+1}\right).
\]
Taking $\gamma=1$ in \eqref{eq:lattice-half-integer-tail} and
$\gamma=-1$ in \eqref{eq:lattice-integer-limit} yields
\[
 \lim_{h\to0^+}\lim_{n\to\infty}\frac{d'(0)}{2T}
 =\frac\pi4\left(0+\frac1{1+1}\right)=\frac\pi8.
\]
Combining these limits, we obtain
\[
 \lim_{h\to0^+}\lim_{n\to\infty}
 \frac{d(\alpha)}{d(1/2)}=\frac{\sin(\pi\alpha)}{1+\cos(\pi\alpha)},\qquad
 \lim_{h\to0^+}\lim_{n\to\infty}
 \frac{d'(0)}{2d(1/2)}=\frac\pi4.
\]

For the KMS case, both boundary values are $a_n$.
For fixed $h$, put $t_k=(k+1/2)T=(k+1/2)\pi/h$ and define the probability weights
\[
 w_{k,h}=\frac{C(t_k)^{-1}}{\sum_{j\in\mathbb Z}C(t_j)^{-1}},
 \qquad \sum_{k\in\mathbb Z}w_{k,h}=1.
\]
As in the general case, the Poisson formulas \eqref{eq:strip-poisson} and
\eqref{eq:poisson-midpoint}, the endpoint identity
\eqref{eq:poisson-endpoint-right} with $b_n$ replaced by $a_n$, and
\eqref{eq:sampling-sine}--\eqref{eq:sampling-singular} give
\[
 \lim_{n\to\infty}\frac{d(1/2)}T
 =\sum_{j\in\mathbb Z}\frac1{C(t_j)}>0.
\]
Since $|t_k|\ge T/2$ and the weights sum to one, the ratio limits satisfy
\begin{equation}\label{eq:kms-interior-sampling}
 \begin{aligned}
 0\le\frac1{\sin(\pi\alpha)}
   \lim_{n\to\infty}\frac{d(\alpha)}{d(1/2)}-1
 &=\sum_{k\in\mathbb Z}w_{k,h}
   \left(\frac{C(t_k)^2}{C(t_k)^2-\cos^2(\pi\alpha)}-1\right)\\
 &\le\frac1{C(T/2)^2-1},
 \end{aligned}
\end{equation}
\begin{equation}\label{eq:kms-endpoint-sampling}
 \begin{aligned}
 0\le\frac2\pi
   \lim_{n\to\infty}\frac{d'(0)}{2d(1/2)}-1
 &=\sum_{k\in\mathbb Z}w_{k,h}
   \left(\frac{C(t_k)^2}{C(t_k)^2-1}-1\right)\\
 &\le\frac1{C(T/2)^2-1}.
 \end{aligned}
\end{equation}
The common upper bound in \eqref{eq:kms-interior-sampling} and
\eqref{eq:kms-endpoint-sampling} tends to zero as $h\to0^+$, so we obtain
\[
 \lim_{h\to0^+}\lim_{n\to\infty}
 \frac{d(\alpha)}{d(1/2)}=\sin(\pi\alpha),\qquad
 \lim_{h\to0^+}\lim_{n\to\infty}
 \frac{d'(0)}{2d(1/2)}=\frac\pi2.
\]
In both classes $d(1/2)>0$ at every finite $h,n$, by its Poisson
integral \eqref{eq:poisson-midpoint} of the nonzero nonnegative
boundary data.

\emph{Strictly positive tests for $1<p<\infty$.}
Fix a finite model and put $\rho_\delta=P+\delta I$, $\delta>0$.
For every fixed $\beta>0$,
\[
 \lim_{\delta\to0^+}\rho_\delta^\beta
 =\lim_{\delta\to0^+}\left\{\delta^\beta I+
       \bigl[(1+\delta)^\beta-\delta^\beta\bigr]P\right\}=P.
\]
For $0<\alpha<1$, substitute $M=\mathcal L$, $A=\rho_\delta$,
and $z=\alpha$ or $z=1/2$ in \eqref{eq:channel-function}, and use
$D_{\rho_\delta}^{\mathcal L}(z)=-F_{\mathcal L,\rho_\delta}(z)$
from \eqref{eq:analytic-energy}. The power limits above give
\[
 \lim_{\delta\to0^+}D_{\rho_\delta}^{\mathcal L}(\alpha)=d(\alpha),
 \qquad
 \lim_{\delta\to0^+}D_{\rho_\delta}^{\mathcal L}(1/2)=d(1/2)>0.
\]
Furthermore, \eqref{eq:generator-reflection} gives
\[
 \lim_{\delta\to0^+}D_{\rho_\delta}^{\mathcal L^\sharp}(\alpha)=d(1-\alpha).
\]
For $1<p<\infty$, set $\alpha=1/p$ and choose the actual generator
\[
 \mathcal A_p=
 \begin{cases}
 \mathcal L^\sharp,&\epsilon=-1\ \text{and }1<p<2,\\
 \mathcal L,&\text{otherwise}.
 \end{cases}
\]
Both choices preserve $\sigma$; in the KMS case they coincide.
Set
\[
 f_{\delta,p}=\Gamma_\sigma^{-1/p}(\rho_\delta^{1/p})>0,
 \qquad
 I_{2,p}f_{\delta,p}=\Gamma_\sigma^{-1/2}(\sqrt{\rho_\delta}).
\]
By \eqref{eq:energy-transport} and \eqref{eq:quadratic-adjoint},
\[
 \lim_{\delta\to0^+}
 \frac{\mathcal E_p^{\mathcal A_p}(f_{\delta,p})}
      {\mathcal E_2^{\mathcal A_p}(I_{2,p}f_{\delta,p})}
 =
 \begin{cases}
 \displaystyle\frac{d(1-\alpha)}{2(1-\alpha)d(1/2)},
       &\epsilon=-1,\quad1<p<2,\\[6pt]
 \displaystyle\frac{d(\alpha)}{2(1-\alpha)d(1/2)},
       &\text{otherwise}.
 \end{cases}
\]
The actual denominators are positive for all sufficiently small
$\delta>0$. The scalar limits give $\kappa_{\mathrm G}(p)$ or
$\kappa_{\mathrm K}(p)$ after the indicated reflection.
At $p=2$, the quotient is exactly one.

\emph{The independent entropy endpoint at $p=1$.}
Start with $\mathcal L$ constructed above, and use
$\mathcal L^\sharp$ as the actual generator at $p=1$.
Since $W\widehat\psi=0$, the trace adjoint satisfies
\[
 \mathcal L_*(P)=-K_0^*P-PK_0,\qquad
 \Tr[P\mathcal L_*(P)]=0.
\]
With $Z=(\log|B|)\otimes I_2$, differentiating
$d(z)=2\langle\psi,|B|^zK|B|^{-z}\psi\rangle$ at $z=0$ gives
\[
 \begin{aligned}
 d'(0)&=4\Ree\langle\widehat\psi,ZK_0\widehat\psi\rangle\\
      &=\Tr[\mathcal L_*(P)\log\sigma].
 \end{aligned}
\]
Here \eqref{eq:model-unital} and $\sum_j\psi_j=0$ give
$K_0^*\widehat\psi=-K_0\widehat\psi$; the scalar part of
$\log\sigma=2Z-\log(2\Tr B^2)I$ drops out by trace preservation.
The two spectral projections of $\rho_\delta$ give
\[
 \log\rho_\delta=(\log\delta)I+
                   \log\frac{1+\delta}{\delta}\,P.
\]
Consequently,
\[
 \lim_{\delta\to0^+}
 \Tr[\mathcal L_*(\rho_\delta)\log\rho_\delta]
 =\lim_{\delta\to0^+}\delta\log\frac{1+\delta}{\delta}
             \Tr[P\mathcal L_*(I)]=0,
\]
and hence
\[
 \begin{aligned}
 \mathcal I_{\mathcal L,\sigma}(\rho_\delta)
 &=-\Tr[\mathcal L_*(\rho_\delta)\log\rho_\delta]
    +\Tr[\mathcal L_*(\rho_\delta)\log\sigma],\\
 \lim_{\delta\to0^+}\mathcal I_{\mathcal L,\sigma}(\rho_\delta)
 &=d'(0).
 \end{aligned}
\]
Apply Proposition~\ref{prop:endpoint} to the generator $\mathcal L^\sharp$
with $f=\Gamma_\sigma^{-1}\rho_\delta$.
For the denominator, take
$X=\Gamma_\sigma^{-1/2}\sqrt{\rho_\delta}=I_{2,1}f$
in \eqref{eq:quadratic-adjoint}, and use the midpoint identity in
\eqref{eq:energy-transport} with $p=1$ and $A_1=\rho_\delta$.
These substitutions give
\[
 \begin{aligned}
 \lim_{\delta\to0^+}\mathcal E_1^{\mathcal L^\sharp}(\Gamma_\sigma^{-1}\rho_\delta)
   &=\tfrac12\lim_{\delta\to0^+}\mathcal I_{\mathcal L,\sigma}(\rho_\delta)
     =\frac{d'(0)}2,\\
 \lim_{\delta\to0^+}\mathcal E_2^{\mathcal L^\sharp}
       (\Gamma_\sigma^{-1/2}\sqrt{\rho_\delta})
   &=\lim_{\delta\to0^+}D_{\rho_\delta}^{\mathcal L}(1/2)=d(1/2)>0.
 \end{aligned}
\]
Thus the endpoint quotient tends to $d'(0)/[2d(1/2)]$, giving
$\pi/4$ in the general case and $\pi/2$ in the KMS case.
This uses the independent entropy calculation, rather than a limit
of the quotients as $p\to1^+$.
\end{proof}

\begin{remark}\label{rem:primitive}
At every finite $p\ge1$, restriction to primitive generators leaves
the optimal constants unchanged in both the general and KMS classes.
Indeed, let $E_\sigma(X)=\Tr(\sigma X)I$. Since
$T_tE_\sigma=E_\sigma T_t=E_\sigma$, the reset perturbation
\begin{equation}\label{eq:primitive-reset}
 T_{\varepsilon,t}=e^{-\varepsilon t}T_t+
                       (1-e^{-\varepsilon t})E_\sigma,
 \qquad\varepsilon>0,
\end{equation}
is a $\sigma$-preserving UCP semigroup with generator
$\mathcal L_\varepsilon=\mathcal L+\varepsilon(E_\sigma-\id)$.
It is primitive, meaning that $T_{\varepsilon,t}\to E_\sigma$ as
$t\to\infty$, and is positivity improving at every $t>0$.
Both KMS and GNS symmetry are preserved when present.
For each fixed $p$ and $f>0$ with
$\mathcal E_2^{\mathcal L}(I_{2,p}f)>0$, both Dirichlet forms are
affine in $\varepsilon$, including at $p=1$ by \eqref{eq:energy-one}.
Thus the denominator stays positive for small $\varepsilon$, and the
quotient converges to its unperturbed value as $\varepsilon\to0^+$.
Every strict violation therefore persists with dimension, state, and
input fixed. Inclusion of the primitive subclass gives the reverse
inequality between the optimal constants.
\end{remark}

\begin{proof}[Proof of Theorem~\ref{thm:optimal-class}]
The result follows from Theorems~\ref{thm:comparison}
and~\ref{thm:sharpness}, together with Remark~\ref{rem:primitive}.
\end{proof}

\appendix

\section{\texorpdfstring{$L_p$}{Lp}-regularity under GNS symmetry}\label{sec:benchmarks}

We record the GNS $L_p$-regularity bound \cite[Theorem 4.1]{Bardet17},
\cite[Theorem 14]{BDR20} in our normalization, extend it to nonprimitive
QMS, and give a matching positive-input witness.

\begin{proposition}\label{prop:benchmarks}
Let $\mathcal H$ be a finite-dimensional Hilbert space and $\mathcal L$ generate a
GNS-symmetric QMS on $B(\mathcal H)$ with faithful invariant state
$\sigma$. For every $1\le p<\infty$ and $f>0$,
\[
 \mathcal E_p^{\mathcal L}(f)
 \ge\frac2p\mathcal E_2^{\mathcal L}(I_{2,p}f).
\]
\end{proposition}

The constant $2/p$ is optimal, even within the primitive
GNS-symmetric class.
\begin{proof}
For primitive GNS-symmetric QMS, \cite[Theorem 14]{BDR20}, whose
$L_p$ Dirichlet form is $p/2$ times ours and whose generator is $-\mathcal L$,
gives the lower bound for $1<p\le2$. GNS implies KMS symmetry
\mbox{\cite[Lemma~12]{BDR20}}; Propositions~\ref{prop:duality}
and~\ref{prop:endpoint} give the other exponents and $p=1$.

The reset approximation in Remark~\ref{rem:primitive}
and continuity of the Dirichlet forms in the generator at each fixed
$f>0$, including at $p=1$, extend the bound to nonprimitive
GNS-symmetric generators.

For optimality, take the reset generator
\[
 \mathcal R(X)=\tfrac12\Tr(X)I_2-X,\qquad
 \sigma=I_2/2,\qquad
 f_\delta=\operatorname{diag}(1+\delta,1-\delta),
 \quad 0<|\delta|<1.
\]
It is primitive and GNS-symmetric, and its $L_2$ Dirichlet form is the
variance for $\Tr/2$. Taylor expansion gives
\[
 \lim_{\delta\to0}
 \frac{\mathcal E_p^{\mathcal R}(f_\delta)}
      {\mathcal E_2^{\mathcal R}(f_\delta^{p/2})}
 =\frac2p,\qquad 1\le p<\infty.
\]
At $p=1$ use the
logarithmic endpoint expansion. This proves the primitive upper bound.
\end{proof}

\bibliographystyle{amsalpha}
\bibliography{regularity}

\indent\textsc{Lijun Wang}\par
\indent\textit{School of Mathematics and Statistics,
Wuhan University, Wuhan 430072, China}\par
\indent\textit{E-mail address: }
\texttt{lijun\_wang@whu.edu.cn}\par
\end{document}